\documentclass[twocolumn,prl,showpacs,amsfonts,amsmath,assymb,eufrak,superscriptaddress]{revtex4-1}
\usepackage{amsmath,amsfonts,amssymb,mathtools,upgreek}
\usepackage{epsfig}

\hypersetup{hidelinks}

\def\eq#1\en{\begin{equation}#1\end{equation}} 
\def\eqa#1\ena{\begin{align}#1\end{align}}
\def\eqg#1\eng{\begin{gather}#1\end{gather}}
\newcommand{\lb}[1]{\label{e:#1}}
\newcommand{\rlb}[1]{\eqref{e:#1}} 
\newcommand{\nl}{\notag\\}

\newcommand{\snorm}[1]{\Vert#1\Vert}

\newcommand{\sbkt}[1]{\langle#1\rangle}

\newcommand{\sumtwo}[2]%
{\mathop{\sum_{#1}}_{#2}}
\newcommand{\sumthree}[3]%
{\mathop{\mathop{\sum_{#1}}_{#2}}_{#3}}
\newcommand{\sumfour}[4]%
{\mathop{\mathop{\mathop{\sum_{#1}}_{#2}}_{#3}}_{#4}} 
\newcommand{\prodtwo}[2]%
{\mathop{\prod_{#1}}_{#2}}
\newcommand{\mintwo}[2]%
{\mathop{\min_{#1}}_{#2}}
\newcommand{\maxtwo}[2]%
{\mathop{\max_{#1}}_{#2}}
\newcommand{\maxthree}[3]%
{\mathop{\mathop{\max_{#1}}_{#2}}_{#3}}
\newcommand{\limtwo}[2]%
{\mathop{\lim_{#1}}_{#2}}
\newcommand{\suptwo}[2]%
{\mathop{\sup_{#1}}_{#2}}
\newcommand{\supthree}[3]%
{\mathop{\mathop{\sup_{#1}}_{#2}}_{#3}}
\newcommand{\supfour}[4]%
{\mathop{\mathop{\mathop{\sup_{#1}}_{#2}}_{#3}}_{#4}} 
\newcommand{\inftwo}[2]%
{\mathop{\inf_{#1}}_{#2}}
\newcommand{\infthree}[3]%
{\mathop{\mathop{\inf_{#1}}_{#2}}_{#3}}
\newcommand{\inffour}[4]%
{\mathop{\mathop{\mathop{\inf_{#1}}_{#2}}_{#3}}_{#4}} 

\newcommand{\mrx}{\mathrm{x}}
\newcommand{\mry}{\mathrm{y}}
\newcommand{\mrz}{\mathrm{z}}

\newcommand{\bbC}{\mathbb{C}}

\newcommand{\bbN}{\mathbb{N}}

\newcommand{\bbR}{\mathbb{R}}
\newcommand{\bbZ}{\mathbb{Z}}

\newcommand{\up}{\uparrow}

\newcommand{\Di}{\mathit{\Delta}}

\newcommand{\qedm}{\rule{1.5mm}{3mm}}

\newcommand{\bra}[1]{\langle#1|}
\newcommand{\ket}[1]{|#1\rangle}

\newcommand{\hA}{\hat{A}}
\newcommand{\hH}{\hat{H}}

\newcommand{\hU}{\hat{U}}
\newcommand{\hV}{\hat{V}}

\newcommand{\hS}{\hat{S}}
\newcommand{\hbS}{\hat{\boldsymbol{S}}}
\newcommand{\hSx}{\hat{S}^\mathrm{x}}
\newcommand{\hSy}{\hat{S}^\mathrm{y}}
\newcommand{\hSz}{\hat{S}^\mathrm{z}}
\newcommand{\hSzt}{\hSz_{\rm tot}}

\newcommand{\EGS}{E^\mathrm{GS}_L}
\newcommand{\Efst}{E^\mathrm{1st}_L}
\newcommand{\GSL}{\ket{\Phi^\mathrm{GS}_L}}
\newcommand{\bGSL}{\bra{\Phi^\mathrm{GS}_L}}

\newcommand{\oo}{\omega^{(1)}}

\newcommand{\Ind}{\text{Ind}}
\newcommand{\Ua}{\hU^{(\alpha)}}
\newcommand{\Usi}{\hU_{\rm si}}
\newcommand{\Ho}{\hH^{(1)}}

\newcommand{\Aloc}{\mathfrak{A}_{\rm loc}}

\newcommand{\para}[1]{\medskip\par{\em #1}\/.---}

\usepackage{color}
\definecolor{fluorescentpink}{rgb}{1.0, 0.08, 0.58}
\definecolor{forestgreen}{rgb}{0.13, 0.55, 0.13}

\newcounter{ct}
\newcommand{\ctl}[1]{\refstepcounter{ct}\thect\label{#1}}

\newcommand{\hSa}{\hat{S}^{(\alpha)}}
\newcommand{\hh}{\hat{h}}
\newcommand{\norms}[1]{\Vert#1\Vert}
\newcommand{\OA}{\mathfrak{A}}
\newcommand{\iop}{\hat{1}}
\newcommand{\GS}{\ket{\Phi^{\rm GS}}}
\newcommand{\GSb}{\bra{\Phi^{\rm GS}}}
\newcommand{\DE}{\Di E}

\begin{document}
\title{The Ground State of the $S=1$ Antiferromagnetic Heisenberg Chain is Topologically Nontrivial if Gapped}

\author{Hal Tasaki}\email[]{hal.tasaki@gakushuin.ac.jp}
\affiliation{Department of Physics, Gakushuin University, Mejiro, Toshima-ku, Tokyo 171-8588, Japan}

\date{\today}

\begin{abstract}
Under the widely accepted but unproven assumption that the one-dimensional $S=1$ antiferromagnetic Heisenberg model has a unique gapped ground state, we prove that the model belongs to a nontrivial symmetry-protected topological (SPT) phase.
In other words, we rigorously rule out the possibility that the model has a unique gapped ground state that is topologically trivial.
To be precise, we assume that the models on open finite chains with boundary magnetic field have unique ground states with a uniform gap and prove that the ground state of the infinite chain has a nontrivial topological index.
This further implies the presence of a gapless edge excitation in the model on the half-infinite chain and the existence of a topological phase transition in the model that interpolates between the Heisenberg chain and the trivial model.

\par\noindent
{\em There is a 23-minute video that explains the motivation and the main results of the present work:} 
 \\\url{https://youtu.be/sq4bNaDJt9g}

\end{abstract}

\maketitle

Haldane's discovery \cite{Haldane1981,Haldane1983a,Haldane1983b,Haldane2017,AKLT1,AKLT2} that the antiferromagnetic Heisenberg chain with integer spin $S$ has a unique gapped ground state opened the rich field of topological phases of matter in quantum spin systems.
It is accepted that the ground state of the model with odd $S$ belongs to a nontrivial symmetry-protected topological (SPT) phase, while that for even $S$ belongs to a trivial SPT phase \cite{Oshikawa1992,GuWen2009,PollmannTurnerBergOshikawa2010,PollmannTurnerBergOshikawa2012,Tasaki2018,Ogata1,Ogata2}.
Since topological phases can never be characterized by a local order parameter, a major challenge in the research was to identify topological indices that characterize the ground states on the infinite chain.
The index defined by Pollmann, Turner, Berg, and Oshikawa for matrix product states \cite{PollmannTurnerBergOshikawa2010,PollmannTurnerBergOshikawa2012} (see also \cite{denNijsRommelse,KennedyTasaki1992A,KennedyTasaki1992B,Perez-Garcia2008,BachmannNachtergaele2014}) was fully generalized by Ogata \cite{Ogata1,Ogata2}.
The evaluation of the indices in the solvable Affleck-Kennedy-Lieb-Tasaki (AKLT) model \cite{AKLT1,AKLT2} rigorously confirmed the above general picture of SPT phases.
See, e.g., Part~II of \cite{TasakiBook} for an overview.

In spite of all the progress, essentially nothing rigorous was established for the most standard antiferromagnetic Heisenberg chain.
The similarity between the Heisenberg and the AKLT models, which seems to be taken for granted, lacks theoretical justification.
Although there has been remarkable progress over the decades in the proof of the existence \cite{Knabe,FannesNachtergaeleWerner1992,Nachtergaele1996,Young2023} and the stability \cite{Yarotsky,BravyiHastingsMichalakis2010,NachtergaeleSimsYoung2022} of the energy gap in many-body quantum systems, none of them are sufficient to show the existence of a gap in the $S=1$ antiferromagnetic Heisenberg chain.

In the present paper, we prove that the ground state of the antiferromagnetic Heisenberg chain with odd $S$ has a nontrivial topological index and hence belongs to a nontrivial SPT phase, under the assumption that the models on open finite chains with boundary field have unique ground states with a uniform energy gap.
In other words, we rigorously establish that the antiferromagnetic Heisenberg chain with odd $S$ cannot have a unique gapped ground state that is topologically trivial.

The proof is based on the elementary index theory developed by Tasaki \cite{Tasaki2018}.
We first use the standard Lieb-Mattis-type technique to evaluate a quantity for a finite chain that corresponds to the index.
The assumption about the existence of a uniform gap then allows us to convert this information to a rigorous evaluation of the index for the ground state of the infinite chain.

We here focus on the model with $S=1$.
All the results readily extend to models with odd $S$, as we shall briefly discuss.
We summarize basic definitions and facts about ground states in the infinite chain in End Matter.

\para{Main results}
The (formal) Hamiltonian of the $S=1$ antiferromagnetic Heisenberg model on the infinite chain is
\eq
\Ho=\sum_{j\in\bbZ}\hbS_j\cdot\hbS_{j+1},
\lb{H}
\en
where $\hbS_j=(\hSx_j,\hSy_j,\hSz_j)$ with $(\hbS_j)^2=2$ denotes the $S=1$ spin operator at site $j$.
Rather than working on the infinite chain directly, we first consider the corresponding model on the finite open chain $\{-L,\ldots,L\}\subset\bbZ$ with the Hamiltonian
\eq
\Ho_L=\sum_{j=-L}^{L-1}\hbS_j\cdot\hbS_{j+1}-h(\hSz_{-L}+\hSz_L),
\lb{HL}
\en
where magnetic field $h>0$ in the positive-z direction is applied to the spins at the two boundary sites.
It should be noted that the Hamiltonian \rlb{HL} does not have any of the three standard symmetries --- the time-reversal symmetry, the $\bbZ_2\times\bbZ_2$ symmetry, or the bond-centered inversion symmetry ---  that protect the Haldane phase \cite{GuWen2009,PollmannTurnerBergOshikawa2010,PollmannTurnerBergOshikawa2012,Tasaki2018,Ogata1,Ogata2}.
Instead, \rlb{HL} is invariant under any uniform spin-rotation about the z-axis and the inversion about the origin.
Fuji, Pollmann, and Oshikawa found that such ${\rm U}(1)\times\bbZ_2$ symmetry also protects the Haldane phase \cite{Fuji}.

By using the standard argument employed to prove the Marshall-Lieb-Mattis theorem \cite{Marshall,LiebMattis1962}, we can prove the following.

\para{Lemma \ctl{LS}}
For any $L=1,2,\ldots$ and $h>0$, the Hamiltonian $\Ho_L$ has a unique ground state $\GSL$.
It satisfies $\hSzt\GSL=\GSL$, where $\hSzt=\sum_{j=-L}^L\hSz_j$.

\para{Proof}
Let us consider the general Hamiltonian
\eq
\hH'_L=\sum_{j=-L}^{L-1}\hbS_j\cdot\hbS_{j+1}-\sum_{j=-L}^L(-1)^{j+L}h_j\hSz_j,
\lb{H'L}
\en
with $h_j\ge0$.
We shall prove the desired statements, assuming $h_j>0$ for at least one $j$.
It is convenient to make a suitable rotation to \rlb{H'L} and treat
\eq
\hH''_L=\sum_{j=-L}^{L-1}\hbS_j\cdot\hbS_{j+1}-\sum_{j=-L}^L(-1)^{j+L}h_j\hSx_j.
\lb{H''L}
\en
With the unitary $\hV=\exp[-i\pi\sum_{j:\text{$j+L$ is odd}}\hSz_j]$, we have
\eqa
\hV^\dagger\hH''_L\hV&=\sum_{j=-L}^{L-1}\{-\tfrac{1}{2}(\hS^+_j\hS^-_{j+1}+\hS^+_j\hS^-_{j+1})+\hSz_j\hSz_{j+1}\}
\nl&-\tfrac{1}{2}\sum_{j=-L}^Lh_j(\hS^+_j+\hS^-_j).
\lb{VHV}
\ena
Let $\{\ket{-}_j,\ket{0}_j,\ket{+}_j\}$ be the basis of the spin at site $j$. 
The standard basis state for the spin chain is $\bigotimes_{j=-L}^L\ket{\sigma_j}_j$, where $(\sigma_{-L},\ldots,\sigma_L)$ with $\sigma_j=0,\pm$ gives a spin configuration.
Then \rlb{VHV} shows that, in the modified basis $\{\hV\bigotimes_{j=-L}^L\ket{\sigma_j}_j\}$, the Hamiltonian $\hH''_L$ has non-positive off-diagonal elements and also all the basis states are connected via nonvanishing matrix elements if $h_j\ne0$ for at least one $j$.
Then, the Perron-Frobenius theorem implies the ground state is unique.
See, e.g., \cite{TasakiBook} for details.

Since $[\hSzt,\Ho_L]=0$, we see $\hSzt\GSL=M\GSL$ for some $M$.
Note that the uniqueness implies $M$ is independent of the magnetic field.
To see $M=1$, it suffices to note that the ground state of \rlb{H'L} with $h_j\gg1$ for all $j$ is essentially the N\'eel state.~\qedm

\medskip

Let $\EGS$ and $\Efst$ be the ground state energy and the first excited energy of $\Ho_L$.
The uniqueness of the ground state implies $\Efst-\EGS>0$.
We shall make the following much stronger assumption, which is nothing but the Haldane conjecture.

\para{Assumption \ctl{AS}}
There are constants $h>0$, $\gamma>0$, and $L_0$ such that
\eq
\Efst-\EGS\ge\gamma,
\lb{gap}
\en
for all $L\ge L_0$ \cite{Assumption}.
\medskip

The assumption is supported by the similarity between the Heisenberg chain and the AKLT chain, which has been repeatedly confirmed numerically.
It is believed that the assumption is valid for a reasonable value of $h$, say $h=1$, with the Haldane gap $\gamma\simeq0.41$.
See, e.g., \cite{TasakiBook}.
As we noted in the introduction, rigorous proof seems far beyond our reach for the moment.

Note that the boundary field in \rlb{HL} is necessary since the model with $h=0$ is believed to possess gapless excitations at the boundaries.
(One can prove that $\Ho_L$ with $h=0$ has three-fold degenerate ground states.)

It should be stressed that we are not sneaking in the topological nature by Assumption~\ref{AS}.
This is most clearly seen by noting that the topologically trivial model with the Hamiltonian
\eq
\hH^{(0)}_L=\sum_{j=-L}^L(\hSz_j)^2-h(\hSz_{-L}+\hSz_L),
\en
also satisfies Assumption~\ref{AS} with $h>0$ such that $h\ne1$ and $\gamma=\min\{|h-1|,1\}$.

Let us define the infinite volume ground state $\oo$ by 
\eq
\oo(\hA)=\lim_{L\up\infty}\bGSL\hA\GSL,
\lb{oo}
\en
where $\GSL$ is the unique normalized ground state of \rlb{HL} and $\hA$ is an arbitrary local operator.
Then, under Assumption~\ref{AS}, we see $\oo$ is a locally-unique gapped ground state of the infinite chain Hamiltonian \rlb{H}.
See End Matter.

Under Assumption~\ref{AS}, we prove the following.

\para{Theorem \ctl{TH}}
The topological index  (defined in \cite{Tasaki2018}) of the ground state $\oo$ is $\Ind[\oo]=-1$.
(See \rlb{Ind} below for the definition of the index.)
Therefore, $\oo$ is in a nontrivial SPT phase.

\medskip
An important implication of the theorem is the existence of a gapless edge excitation in the model on the half-infinite chain $\bbZ_+=\{0,1,\ldots\}$ \cite{AKLT2,Kennedy1990,Hagiwaraetal1990}.
See End Matter for the definition of corresponding ground states.
Let $\oo_+$ be a ground state of $\Ho_+=\sum_{j\in\bbZ_+}\hbS_j\cdot\hbS_{j+1}$ that is invariant under the uniform $\pi$-rotation about the x-axis.

\para{Corollary \ctl{EDGE}}
We additionally assume that the ground state of \rlb{H} is unique.
Then, for any $\varepsilon>0$, there is a unitary operator $\hV_\varepsilon$ acting only on a finite number of sites such that $\oo_+(\hV_\varepsilon)=0$ and $\oo_+(\hV_\varepsilon^\dagger\Ho_+\hV_\varepsilon-\Ho_+)\le\varepsilon$.

\medskip
In other words, $\hV_\varepsilon$ generates a state orthogonal to the ground state $\oo_+$ with excitation energy not exceeding $\varepsilon$.
A key for the proof is that the unique ground state $\oo$ has the same symmetry as the Hamiltonian \rlb{H}.
This allows us to define $\bbZ_2$ topological indices corresponding to the three classes of symmetry discussed in \cite{Tasaki2018}.
Corollary~\ref{EDGE} is proved in exactly the same way as Theorem~3.6 of \cite{Tasaki2023} for fermion models by using the symmetry with respect to the $\pi$-rotation about the x-axis.
See also \cite{Tasaki_Lieb90,TasakiNext}.

Theorem~\ref{TH} also  implies the existence of a topological phase transition in the ground states of the one-parameter family of Hamiltonians
\eq
\hH^{(s)}=\sum_{j\in\bbZ}\{s\,\hbS_j\cdot\hbS_{j+1}+(1-s)(\hSz_j)^2\},
\lb{Hs}
\en
where $s\in[0,1]$.
For each $s\in[0,1]$ we let $\omega^{(s)}$ be a ground state of $\hH^{(s)}$.
(One may choose any ground state if the ground states are not unique.)

\para{Corollary \ctl{TP}}
There exists $s_0\in(0,1)$ such that one has either (i)~$\omega^{(s_0)}$ is not a locally-unique gapped ground state, (ii)~the energy gap approaches zero as $s\to s_0$, (iii)~$\omega^{(s_0)}$ (spontaneously) breaks the site-centered inversion symmetry (see below), or (iv)~$\omega^{(s)}(\hA)$ is discontinuous in $s$ at $s_0$ for some local operator $\hA$.

\medskip

We shall discuss the proof in the next part.

In case Assumption~\ref{AS} is not valid, one may regard $\oo$ itself as critical.
Therefore, we have established without any unproven assumptions that the model \rlb{Hs} undergoes a phase transition at some $s_0\in(0,1]$.

\para{The twist operator and the $\bbZ_2$ index}
The definition of the $\bbZ_2$-index in \cite{Tasaki2018} is based on the observation by Nakamura and Todo \cite{NakamuraTodo2002} that the expectation value of the twist operator plays the role of an order parameter for the Haldane phase.
We define the rotation angle by
\eq
\theta^{(\alpha)}_j=
\begin{cases}
0,&j\le-\tfrac{\pi}{\alpha};\\
\pi+\alpha j,&-\tfrac{\pi}{\alpha}\le j\le\tfrac{\pi}{\alpha};\\
2\pi,&j\ge\tfrac{\pi}{\alpha},
\end{cases}
\lb{theta}
\en
for $\alpha>0$ and $\theta_j^{(0)}=\pi$.
We then define the twist operator \cite{Bohm,LSM,AL} on the finite chain $\{-L,\ldots,L\}$ by
\eq
\Ua_L=\exp\Bigl[-i\sum_{j=-L}^L\theta^{(\alpha)}_j\hSz_j\Bigr],
\lb{U1}
\en
for $\alpha\ge0$, and that on the infinite chain by
\eq
\Ua=\exp\Bigl[-i\sum_{j\in\bbZ\cap[-\tfrac{\pi}{\alpha},\tfrac{\pi}{\alpha}]}\theta^{(\alpha)}_j\hSz_j\Bigr],
\lb{U2}
\en
for $\alpha>0$, where $\Ua$ is a local operator.
Also note that \rlb{U2} is obtained as the $L\up\infty$ limit of \rlb{U1} because $e^{-i2\pi\hSz_j}=\hat{1}$.
It is crucial that $\Ua_L$ and $\Ua$ are continuous in $\alpha$.

Let us review the index theory in \cite{Tasaki2018}, restricting our attention to the class of models with the ${\rm U}(1)\times\bbZ_2$ symmetry \cite{symmetry}.
Let $\hH$ be a Hamiltonian for the $S=1$ spin system on the infinite chain $\bbZ$ that is invariant under any uniform rotation about the z-axis.
We assume that $\hH$ has a locally-unique gapped ground state $\omega$ with gap $\gamma'>0$.
We further assume that $\omega$ is invariant under site-centered inversion, i.e., $\omega(\Usi\hA\Usi)=\omega(\hA)$ for any local operator $\hA$.
Here we defined the unitary $\Usi=\Usi^\dagger$ for site-centered inversion by $\Usi\bigotimes_{j=-L}^L\ket{\sigma_j}_j=\bigotimes_{j=-L}^L\ket{\sigma_{-j}}_j$ for any spin configuration $(\sigma_{-L},\ldots,\sigma_L)$.
(We take $L$ large enough so that the support of $\hA$ is included in $\{-L,\ldots,L\}$.)

Then the $\bbZ_2$ index for $\omega$ is defined as
\eq
\Ind[\omega]=\lim_{\alpha\downarrow0}\omega(\Ua)\in\{-1,1\},
\lb{Ind}
\en
where the existence of the limit is guaranteed \cite{Tasaki2018,TasakiNext}.
It is also found that the index $\Ind(\omega)$ is invariant when the Hamiltonian and the ground states are continuously modified so that the above symmetries are preserved and the ground state remains locally-unique and gapped.

To prove Corollary~\ref{TP}, we first note $\Ind[\omega^{(0)}]=1$.
This is becuase the ground state of $\hH^{(0)}$ is $\bigotimes_{j\in\bbZ}\ket{0}_j$, and hence
$\omega^{(0)}(\Ua)=1$ for any $\alpha>0$.
Assume that $\omega^{(s)}$ does not undergo a phase transition, i.e., $\omega^{(s)}$ is a locally-unique gapped ground state with a uniform gap, preserves the site-centered inversion symmetry, and is continuous in $s$.
Then, the invariance of the index implies $\Ind[\oo]=1$, which contradicts Theorem~\ref{TH}.

\para{Proof of Theorem~\ref{TH}}
The following simple lemma is essential.
\para{Lemma \ctl{RE}}
$\bGSL\Ua_L\GSL\in\bbR$.

\para{Proof}
In this proof, we abbreviate $\bGSL\cdot\GSL$ as $\sbkt{\cdot}$.
Let $\Usi$ be the unitary for site-centered inversion on the chain $\{-L,\ldots,L\}$ defined as above.
Since $[\Usi,\Ho_L]=0$, the uniqueness implies $\Usi\GSL=\pm\GSL$ (where we know from the Perron-Frobenius theorem that the sign is $+$) and hence $\sbkt{\hA}=\sbkt{\Usi\hA\Usi}$ for any $\hA$.
We then find
\eqa
&\sbkt{\Ua_L}=\sbkt{\Usi\Ua_L\Usi}=\Bigl\langle\exp\Bigl[-i\sum_{j=-L}^L\theta^{(\alpha)}_{-j}\hSz_j\Bigr]\Bigr\rangle
\nl
&=\Bigl\langle\exp\Bigl[-i\sum_{j=-L}^L(\theta^{(\alpha)}_{-j}-2\pi)\hSz_j\Bigr]\Bigr\rangle
\nl&=\Bigl\langle\exp\Bigl[i\sum_{j=-L}^L\theta^{(\alpha)}_{j}\hSz_j\Bigr]\Bigr\rangle
=\sbkt{(\Ua_L)^\dagger}
=\sbkt{\Ua_L}^*,
\ena
where we noted $\theta^{(\alpha)}_{-j}-2\pi=-\theta^{(\alpha)}_j$.~\qedm

\medskip
We next recall the variational estimate due to Lieb, Schultz, and Mattis \cite{LSM}.
See also \cite{TasakiBook, TasakiLSM} for proof.
\para{Lemma \ctl{LSM}}
For any $\alpha\ge0$ and $L$, one has 
\eq
\bGSL(\Ua_L)^\dagger\Ho_L\Ua_L\GSL-\EGS\le4\alpha^2(\tfrac{\pi}{\alpha}+1).
\lb{LSM}
\en

\medskip
The left-hand side represents the increase in the energy (compared with the ground state energy) in the variational state $\Ua_L\GSL$.
The upper bound is obtained by noting that the extra energy in each twisted bond does not exceed $2\alpha^2$, and there are at most $2(\tfrac{\pi}{\alpha}+1)$ bonds that are twisted.
The bound \rlb{LSM} shows that the energy increase is small for small $\alpha$.

So far, we have not used Assumption~\ref{AS}.
The final lemma relies on the assumption.
\para{Lemma \ctl{BD}}
Let $c(\alpha)=\sqrt{1-\frac{4}{\gamma}(\pi+\alpha)\alpha}$ and $\alpha_0$ be the unique positive root of $c(\alpha_0)=0$.
It is crucial that $c(\alpha)\in(0,1]$ for $\alpha\in[0,\alpha_0)$ and that $c(\alpha)\up1$ as $\alpha\downarrow0$.
For any $\alpha\in[0,\alpha_0)$ and $L\ge L_0$, we have
$|\bGSL\Ua_L\GSL|\ge c(\alpha)$.
\para{Proof}
Let $\ket{\Psi_j}$ with $j=0,1,\ldots$ be the normalized eigenstates of $\Ho_L$ with energy eigenvalue $E_j$, where $j=0$ and 1 correspond to the ground state and the first excited state, respectively.
Expanding the trial state as $\Ua_L\GSL=\sum_jc_j\ket{\Psi_j}$, we find
\eqa
\bGSL(\Ua_L)^\dagger\Ho_L\Ua_L&\GSL-\EGS=\sum_j|c_j|^2(E_j-E_0)
\nl&\ge\sum_{j\ne0}|c_j|^2\gamma=(1-|c_0|^2)\gamma,
\ena
where we noted that \rlb{gap} implies $E_j-E_0\ge\gamma$ for $j\ne0$.
Combining this with \rlb{LSM}, we find $|c_0|^2\ge\{c(\alpha)\}^2$.
Recalling $c_0=\bGSL\Ua_L\GSL$, we get the desired result.~\qedm

\medskip
Note that Lemmas~\ref{RE} and \ref{BD} show that the expectation value $\bGSL\Ua_L\GSL$ is either in $[-1,-c(\alpha)]$ or $[c(\alpha),1]$ (where $c(\alpha)>0$) if $L\ge L_0$ and $\alpha\in[0,\alpha_0)$.
This dichotomy is at the heart of the definition of the $\bbZ_2$ index in \cite{Tasaki2018}.

We are now ready to determine the topological index of the ground state $\oo$.
We first note that for $\alpha=0$, Lemma~\ref{LS} implies
\eq
\bGSL\hU_L^{(0)}\GSL=\bGSL\exp[-i\pi\hSzt]\GSL=-1.
\lb{-1}
\en
Since $\bGSL\Ua\GSL$ (for fixed $L\ge L_0$) is continuous in $\alpha$, the foregoing observation shows $\bGSL\Ua_L\GSL\in[-1,-c(\alpha)]$ for $\alpha\in(0,\alpha_0)$.
Since this holds for any $L\ge L_0$, we see for the infinite volume ground state \rlb{oo} that
\eq
\oo(\Ua)\in[-1,-c(\alpha)]\quad\text{for any $\alpha\in(0,\alpha_0)$,}
\lb{oU}
\en
which, with \rlb{Ind}, shows $\Ind[\oo]=-1$.

\para{General integer $S$}
We can treat the model with general integral spin $S$ with the same Hamiltonians \rlb{H} and \rlb{HL}.
Lemmna~\ref{LS} is still valid with the generalized relation $\hSzt\GSL=S\GSL$.
Corresponding to \rlb{-1}, we then have $\bGSL\hU_L^{(0)}\GSL=e^{-i\pi S}$.
Under Assumption~\ref{AS}, we see $\Ind[\oo]=(-1)^S$.
We conclude the model belongs to a nontrivial SPT phase if $S$ is odd.
The result for even $S$ is consistent with the conjecture that the model belongs to the trivial SPT phase \cite{Oshikawa1992,PollmannTurnerBergOshikawa2010,PollmannTurnerBergOshikawa2012}.

\para{Discussion}
We proved that the antiferromagnetic Heisenberg chain with $S=1$ (or, more generally, odd $S$) belongs to a nontrivial SPT phase, under Assumption~\ref{AS} about the existence of the Haldane gap.
The theorem implies the presence of a gapless edge excitation in the model on the half-infinite chain and the existence of a phase transition in the interpolating model \rlb{Hs}.

Generally speaking, a quantum many-body system may have (i)~a unique gapped ground state that is topologically trivial, (ii)~a unique gapped ground state that is topologically nontrivial, or (iii)~gapless or degenerate ground states.
Our theorem rigorously rules out the (least interesting) possibility (i) for the antiferromagnetic Heisenberg chain with odd $S$.
We have thus established that the ground state of the antiferromagnetic Heisenberg chain with odd $S$ must be nontrivial in some sense.

In our proof, it is essential to treat a model on a finite open chain that has (or is expected to have) a unique gapped ground state and possesses sufficient symmetry to protect the Haldane phase.
The model \rlb{HL} with the ${\rm U}(1)\times\bbZ_2$ symmetry turns out to be essentially the unique choice \cite{S/2}.
Then, an elementary Lieb-Mattis-type argument for a finite chain yields the key identity \rlb{-1}.
Of course, such an identity for finite chains alone is insufficient to determine the topological index for the ground state of the infinite chain.
Here, Lemmas~\ref{RE} and \ref{BD} guarantee that one can continuously increase $\alpha$ (up to $\alpha_0$) and then take the $L\up\infty$ limit.

The whole analysis makes use of the index defined in \cite{Tasaki2018} in terms of the expectation value of the twist operator.
We believe that the index coincides with the Ogata index \cite{Ogata1,Ogata2} for the same model.
However, proving the two indices are identical is a nontrivial problem since the Ogata index is defined in terms of a projective representation of the symmetry group defined on the infinite chain and is not easy to evaluate in general.

After having proven that ``the ground state of the $S=1$ antiferromagnetic Heisenberg chain is topologically nontrivial if gapped'', the remaining major challenge is to prove the existence of a gap.
The currently available techniques for rigorously controlling the gap of many-body quantum systems \cite{Knabe,FannesNachtergaeleWerner1992,Nachtergaele1996,Young2023,Yarotsky,BravyiHastingsMichalakis2010,NachtergaeleSimsYoung2022} apply only to frustration-free models (and their perturbations) and are insufficient to justify Assumption~\ref{AS}.
It is challenging to develop a new strategy that would allow computer-aided proof of the existence of a gap in the $S=1$ antiferromagnetic Heisenberg chain.

\medskip
{\noindent
It is a pleasure to thank Hosho Katsura and Akinori Tanaka for their valuable discussions over the years, which made the present work possible, and Sven Bachmann, Yohei Fuji, Tohru Koma, Elliott Lieb, Bruno Nachtergaele, Yoshiko Ogata, Masaki Oshikawa, and Amanda Young for their useful discussions.
I also thank the anonymous referees for their valuable comments.
The present research is supported by JSPS Grants-in-Aid for Scientific Research No. 22K03474.
}


\appendix
\section*{End Matter}

Here, we summarize basic definitions and facts about ground states of quantum spin systems on the infinite chain $\bbZ$.
The discussion applies to any $S=\frac{1}{2},1,\ldots$.
See \cite{TasakiLSM} for further details, generalizations, proofs, and references.

We define the quantum spin system on $\bbZ$ by associating with each site $j\in\bbZ$ the single-spin Hilbert space $\mathfrak{h}_j\cong\bbC^{2S+1}$.
The Hamiltonian is formally expressed as the infinite sum
\eq
\hH=\sum_{j\in\bbZ}\hh_j,
\lb{HZ}
\en
where the local Hamiltonian $\hh_j$ is a polynomial of spin operators $\hSa_k$ with $\alpha=\mrx$, $\mry$, $\mrz$ and $k\in\{j,j+1\}$.
We also assume that the operator norm is bounded as  $\norms{\hh_j}\le h_0$ for any $j\in\bbZ$ with a constant $h_0$.
An important example is \rlb{Hs}.

To define ground states of \rlb{HZ}, it is standard and convenient to work with the algebra of operators rather than with the Hilbert space.
By a local operator of the spin chain, we mean an arbitrary polynomial of operators $\hSa_j$ with $\alpha=\mrx$, $\mry$, $\mrz$ and $j\in\bbZ$.
The set of sites on which a local operator acts nontrivially is called the support of the operator.
Note that the support of a local operator is always a finite subset of $\bbZ$.
We define the algebra of local operators, which is denoted as $\Aloc$, as the set of all local operators.
It is worth noting that the local Hamiltonian $\hh_j$ belongs to $\Aloc$, but the total Hamiltonian $\hH$ does not.

We can then define the notion of states on the infinite chain as follows.
The idea is that $\rho(\hA)$ is the expectation value of the operator $\hA$ in the state.

\para{Definition \ctl{state}}
A state $\rho$ of the spin chain is a liner map from $\Aloc$ to $\bbC$ such that $\rho(\iop)=1$ and $\rho(\hA^\dagger\hA)\ge0$ for any $\hA\in\Aloc$.

\medskip
For an arbitrary local operator $\hV\in\Aloc$, we define its commutator with $\hH$ as
\eq
[\hH,\hV]=\sum_{j=-\ell}^\ell[\hh_j,\hV],
\lb{HV}
\en
where $\ell$ is taken so that the support of $\hV$ is included in $\{-\ell-1,\ldots,\ell\}$.
Note that the definition is independent of the choice of $\ell$.
It is notable that $[\hH,\hV]$ is a well-defined local operator, although $\hH$ is only formally defined by the infinite sum \rlb{H}.
Then a ground state is characterized as follows.

\para{Definition \ctl{GS}}
A state $\omega$ is said to be a ground state of $\hH$ if it holds for any $\hV\in\Aloc$ that 
\eq
\omega(\hV^\dagger[\hH,\hV])\ge0.
\lb{GS1}
\en

\medskip
In short, the definition says that one can never lower the energy expectation value of the state $\omega$ by perturbing it with a local operator $\hV$.
To see this more clearly, consider a finite system and let $\omega$ be a state written as $\omega(\cdot)=\bra{\Phi}\cdot\ket{\Phi}$ with a pure normalized state $\ket{\Phi}$.
Then the condition \rlb{GS1} reads 
\eq
\bra{\Phi}\hV^\dagger\hH\hV\ket{\Phi}\ge\bra{\Phi}\hV^\dagger\hV\hH\ket{\Phi}.
\lb{GS1B}
\en
If we take $\ket{\Phi}$ as a ground state $\GS$ with energy $E_{\rm GS}$, \rlb{GS1B} is rewritten as $\bra{\Psi}\hH\ket{\Psi}\ge E_{\rm GS}$, 
where 
\eq
\ket{\Psi}=\frac{\hV\GS}{\snorm{\hV\GS}}
\lb{Psi}
\en
is a normalized variational state.
We thus get the standard variational characterization of a ground state.
When $\ket{\Phi}$ is not a ground state, one immediately sees \rlb{GS1B} is violated by taking $\hV=\GS\bra{\Phi}$.

In Corollary~\ref{EDGE}, we assumed that the ground state of \rlb{H} is unique.
This means that there is only one state $\omega$ that satisfies the condition in Definition~\ref{GS}.

The above definitions can be extended to spin models on the half-infinite chain $\bbZ_+=\{0,1,\ldots\}$.
One only needs to define the algebra of local operators $\OA_{\rm loc,+}$ on $\bbZ_+$ and corresponding states, and then rewrite Definition~\ref{GS} with a suitable Hamiltonian $\hH_+$ defined on $\bbZ_+$.
The ground state $\oo_+$ treated in Corollary~\ref{EDGE} is defined in this manner.

Let us return to the models on the full infinite chain $\bbZ$ and discuss the notion of a locally-unique gapped ground state.

\para{Definition \ctl{LUGGS}}
A ground state $\omega$ of $\hH$ is said to be a locally-unique gapped ground state if there is a constant $\gamma>0$ such that
\eq
\omega(\hV^\dagger[\hH,\hV])\ge\gamma\,\omega(\hV^\dagger\hV),
\lb{GS2}
\en
holds for any $\hV\in\Aloc$ with $\omega(\hV)=0$.
The energy gap $\DE$ of the ground state $\omega$ is the largest $\gamma$ with the above property.

\medskip
To see the relation with the standard definition for a finite system, observe that the conditions read $\bra{\Psi}\hH\ket{\Psi}\ge E_{\rm GS}+\gamma$ and $\GSb\Psi\rangle=0$ for the same $\GS$ and $\ket{\Psi}$ as above.
These are precisely the variational characterization of a unique gapped ground state.

As the terminology suggests, a locally-unique gapped ground state may not be a unique ground state.

Let us see how the above rather abstract definition of a locally-unique gapped ground state is related to the definition that is standard in physics.
Let $L$ be a positive integer, and consider the quantum spin system on the finite chain $\{-L,\ldots,L\}\subset\bbZ$ with the Hamiltonian
\eq
\hH_L=\sum_{j=-L}^{L-1}\hh_j+\Di\hH_L,
\en
where the local Hamiltonian $\hh_j$ is the same as in \rlb{HZ}.
Here $\Di\hH_L$ is a suitably chosen boundary Hamiltonian that acts on sites within a fixed distance (independent of $L$) from the two boundaries.
As a special case, one can realize the periodic boundary condition.

We assume that, for each $L$, the finite volume Hamiltonian $\hH_L$ has a unique normalized ground state $\GSL$ accompanied by a nonzero energy gap that is not less than a constant $\DE>0$.

We then define a state $\omega$ on the infinite chain by
\eq
\omega(\hA)=\lim_{L\up\infty}\bra{\Phi^{\rm GS}_L}\hA\GSL,
\lb{lim1}
\en
for any $\hA\in\Aloc$.  
Note that the expectation value $\bra{\Phi^{\rm GS}_{L}}\hA\GSL$ is well-defined for sufficiently large $L$ since $\hA$ is local.
Of course, the limit \rlb{lim1} may not exist.
It is known, however, that one can always take a subsequence, i.e., a strictly increasing function $L(n)\in\bbN$ of $n\in\bbN$, such that the infinite volume limit
\eq
\omega(\hA)=\lim_{n\up\infty}\bra{\Phi^{\rm GS}_{L(n)}}\hA\ket{\Phi^{\rm GS}_{L(n)}},
\lb{lim2}
\en
exists for any $\hA\in\Aloc$.
This defines a state of the infinite chain, although the limit may not be unique in general.

\para{Theorem \ctl{finiteGS}}
\label{t:finiteGS}
The limiting infinite volume state $\omega$ is a locally-unique gapped ground state of $\hH$.

\medskip

This important theorem states that if one has a sequence of finite chains with a unique gapped ground state, then the corresponding state in the infinite volume limit is necessarily a locally-unique gapped ground state.
We see that the notion of a locally-unique gapped ground state is natural and useful from the physical point of view.


\begin{thebibliography}{10}
\bibitem{Assumption}
It is sufficient that \rlb{gap} is valid for an infinite increasing sequence of $L$.  The field $h>0$ may depend on $L$.
\bibitem{S/2}
One can also treat a spin $S$ chain without a magnetic field with two spin $S/2$ at the two ends to get the same conclusion \cite{TasakiNext}.  However, assuming that such a model has a unique gapped ground state is close to assuming that the ground state has an AKLT-like structure.
\end{thebibliography}
\end{document}